# Plasmon-Exciton Coupling Effect on Plasmon Damping

Lulu Ye,[1] Weidong Zhang,[1] Aiqin Hu,[1] Hai Lin,[1] Jinglin Tang,[1] Yunkun Wang[1], Chenxinyu Pan, [2] Pan Wang, [2] Xin Guo,[2] Limin Tong,[2] Yunan Gao,[1,3,4] Qihuang Gong,[1,3,4] and Guowei Lu[1,3,4,*]

[1] State Key Laboratory for Mesoscopic Physics, Frontiers Science Center for Nano-optoelectronics & Collaborative Innovation Center of Quantum Matter, School of Physics, Peking University, Beijing 100871, China.

[2] State Key Laboratory of Modern Optical Instrumentation, College of Optical Science and Engineering, Zhejiang University, Hangzhou 310027, China.

[3] Collaborative Innovation Center of Extreme Optics, Shanxi University, Taiyuan, Shanxi 030006, China.

[4] Peking University Yangtze Delta Institute of Optoelectronics, Nantong 226010, Jiangsu, China.

## ABSTRACT

Plasmon decay via the surface or interface is a critical process for practical energy conversion and plasmonic catalysis. However, the relationship between plasmon damping and the coupling between the plasmon and 2D materials is still unclear. The spectral splitting due to plasmon-exciton interaction impedes the conventional single-particle method to evaluate the plasmon damping rate by the spectral linewidth directly. Here, we investigated the interaction between a single gold nanorod (GNR) and 2D materials using the single-particle spectroscopy method assisted with *in situ* nanomanipulation technique by comparing scattering intensity and linewidth together. Our approach allows us to indisputably identify that the plasmon-exciton coupling in the GNR-WSe$_2$ hybrid would induce plasmon damping. We can also isolate the contribution between the charge transfer channel and resonant energy transfer channel for the plasmon decay in the GNR-graphene hybrid by comparing that with thin *h*BN layers as an intermediate medium to block the charge transfer. We find out that the contact layer between the GNR and 2D materials contributes most of the interfacial plasmon damping. These findings contribute to a deep understanding of interfacial excitonic effects on the plasmon and 2D materials hybrid.

**INTRODUCTION**

Plasmonic nanostructures of noble metals have attracted immense interest due to their unique properties as optical antennas, local heat sources, and hot carrier generators[1,2,3]. The localized surface plasmon resonance (LSPR) originating from the light-induced collective oscillation of charge carriers provides an efficient platform to manipulate photons at the nanoscale, exhibiting a huge localized electromagnetic field concentration and large resonant extinction cross-sections. After the initial excitation of an LSPR mode, the coherent electron oscillation decays via several channels: interband or intraband transitions, radiation, and surface damping[4-5]. Because of the different plasmon damping mechanisms, total damping rate depends on the nanostructures' shape, size, material, and surrounding environment, etc. Understanding the various contributions to plasmon damping allows for optimizing the performances of the plasmonic devices, such as applications in sensors, photocatalysis, photodetectors, and plasmon-driven chemistry[6, 7-10].

Plasmon oscillations can decay through interband, intraband, radiation, and surface damping. Bulk damping $\Gamma_b$ is due to electron scattering within the metal, including interband and intraband transitions, and the material's complex dielectric function well describes it. Radiation damping $\Gamma_{rad}$ describes the coupling of the plasmon oscillation to the radiation field, which increases as a function of particle volume. Surface damping $\Gamma_{surf}$ includes a scattering of the electrons at the interface[11], and chemical interface damping (CID) [7, 12-13] refers to different chemical environments. $\Gamma_{CID}$ is attributed to the decay of the coherent oscillation through charge transfer, dipole scattering, resonance energy transfer, etc[14-15]. The CID in catalytic metal nanoislands on gold nanorods has been demonstrated as a major component of the total damping rate of the plasmon resonance[11].

Investigation of plasmon damping at the single-nanoparticle level provides an excellent way to reveal coherent plasmon-exciton interactions[16]. The homogeneous linewidth obtained with the single-particle spectroscopy method has been used when investigating the CID of single gold nanorods (GNRs). For instance, the interfacial interaction due to the adsorbates, such as molecules with thiol sulfur, nano-metal, and graphene, would result in the LSPR spectral frequency red-shift, intensity loss, and the full line width at half-maximum (FWHM) broadening[14, 17]. However, despite these recent advances, two essential aspects of the surface damping investigations have not been resolved so far: (1) the effect of plasmon-exciton coupling on plasmon damping and (2) distinguish the contributions between charge transfer channels and resonant energy transfer channel for the plasmon decay. Therefore, it is necessary to understand better interfacial excitonic effects on the plasmon and 2D materials.

In this study, we investigate the interaction between single GNRs and 2D materials such as $WSe_2$, graphene, and hexagonal boron nitride (hBN) using the single-particle spectroscopy method assisted by *in situ* nanomanipulation via atomic force microscopy (AFM). In the experiment, we measure and compare the scattering spectra of the same individual GNRs before and after interacting with the 2D materials. We find that the plasmon-exciton coupling would result in the scattering intensity decrease. Meanwhile, there is a spectral splitting owing to plasmon-exciton interaction, which means that the plasmon-exciton interaction would cause additional plasmon damping. Furthermore, by comparing GNR/hBN/graphene with GNR/graphene, we verify that the contribution of resonant energy transfer dominates the plasmon damping rather than the electron transfer mechanism in the interaction of the GNR and graphene. We also demonstrate the influence of plasmon-exciton coupling strength and the exciton transition dipole moment on the interfacial interactions by comparing GNR/hBN/$WSe_2$ heterostructure and various layered $WSe_2$ sheets.

**RESULTS AND DISCUSSION**

The scattering spectrum of the metal nanoparticle and 2D materials hybrid system is affected by different interaction processes (Figure 1a). In *situ* single-particle spectroscopy method assisted with the AFM nanomanipulation (Figure 1b) enables us to decouple several factors to reveal the plasmon-exciton interaction effect on the plasmon damping. The optical setup shown in Figure S1 has been described in detail previously[18-19]. Individual GNR can be confirmed by scanning optical confocal images and AFM images. We can manipulate a single GNR using the AFM nanomanipulation method and compare the spectra before and after coupling *in situ* for the same single GNR. The scattering spectra of different particles can be calibrated with a reference nanoparticle on the same sample.

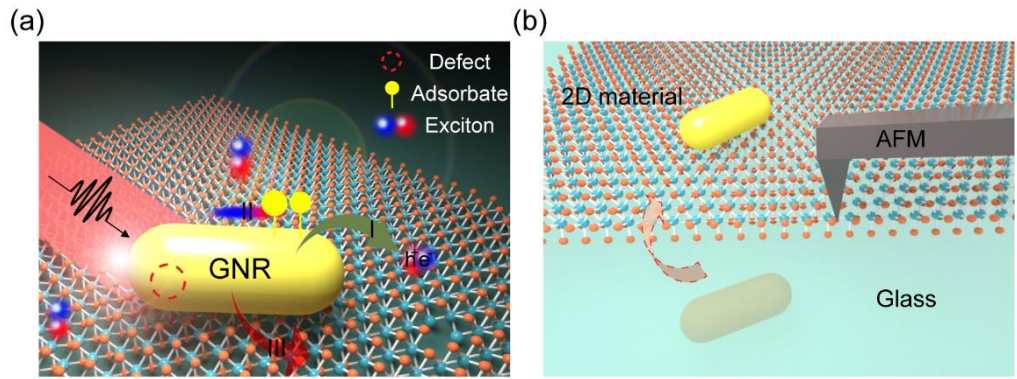

**Figure 1.** Scheme of the GNR-2D materials interactions and experimental configuration. (a) Schematic of plasmon damping: charge transfer, dipole scattering, and resonance energy transfer, surface adsorbates, defect induced dipoles. (b) Schematic of the experiment: a GNR being moved from the glass onto a monolayer WSe$_2$ through the AFM manipulation.

Most of the previous results show that the scattering intensity becomes weaker due to the CID effect[11]. Interestingly, the interaction between a single GNR and monolayer Transition Metal Dichalcogenides (TMDCs) hybrids exhibits an increase in the scattering intensity of the hybrid compared with that before coupling[20]. Here, we investigated the GNRs with the different LSPR frequencies due to different aspect ratios, which are coupled with the monolayer WSe$_2$, to reveal the plasmon-exciton coupling effect. The photoluminescence (PL) spectra of monolayer WSe$_2$ on the glass substrate in air exhibit fluorescent peaks at a wavelength of 752 nm with the FWHM of ~ 51 meV (Figure S2). Figure 2 shows the representative scattering spectra of the same GNRs before and after coupling with the WSe$_2$ monolayer. After coupling, the scattering peak redshifted because of the high refractive index of the TMDCs[21]. Thus, there is a spectral splitting after coupling, and the scattering intensity is dependent on the detuning between exciton and plasmon. Then, variation of the FWHM cannot be applied to evaluate the plasmon damping rate due to the plasmon-exciton coupling[17]. Here, by comparing the variation of scattering intensity of the same particle, we provide another indicator to assess the plasmon damping. As shown in Figure 2, the scattering intensity decreases when the LSPR and exciton are almost resonant (slight detuning) compared to when the detuning is large, although the integral scattering intensity increases compared to that on the glass surface. In a coupling system, the large detuning results in a weak plasmon-exciton coupling strength, which would lead to less decay rate for the plasmon resonance. The results reveal that the plasmon-exciton coupling is an efficient plasmon damping channel.

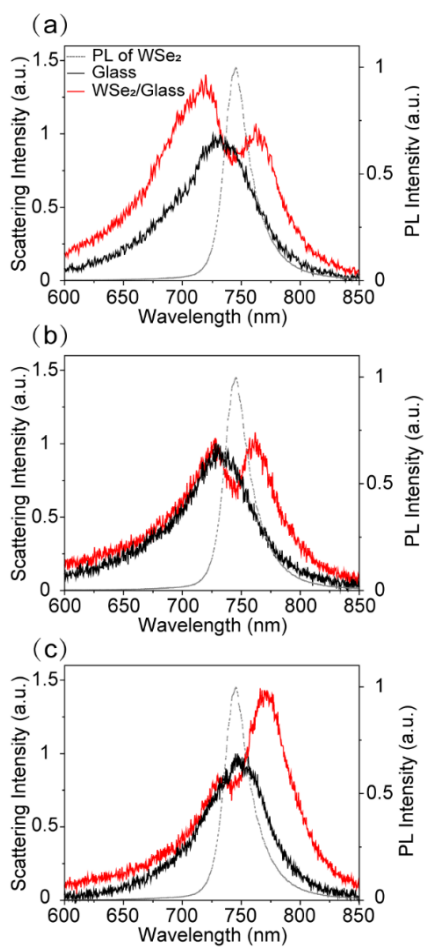

**Figure 2.** The scattering spectra of single GNRs with different detuning with respective the WSe$_2$ exciton, blue detuning (a), small detuning (b), and red detuning (c) by using GNRs with different aspect ratios. Scattering spectra when the same GNR on the glass (black) or the WSe₂ (red) correspondingly, and the PL spectrum of WSe₂ (gray) for comparison.

To further demonstrate the plasmon-exciton coupling effect, we perform single-particle spectroscopy on the layered structure sample consist of $h$BN flakes and monolayer graphene, as shown in Figure 3a. The $h$BN flakes are a layered insulator with a wide bandgap, and their exciton frequency does not overlap the GNR's LSPR anymore. Figure 3b shows the representative scattering spectra before and after the GNR being moved onto the $h$BN. We can see that the scattering peak redshifts ~ 5±1 nm, and the scattering intensity of the same GNR increases compared to that on the glass. The spectral shape on the $h$BN shows a Lorentz line shape with almost the same FWHM as that on the glass. In contrast, the gapless graphene would couple with the GNR's plasmon in broadband. The GNR-graphene hybrid results in the decrease of scattering intensity and the FWHM broadening of the GNR, as shown in Figure 3c[22]. These results confirm the above conclusion that the plasmon-exciton coupling would induce the plasmon damping. Generally, the hybrid system couples with the external electromagnetic field mainly through plasmon resonances. Then, the near-field coupling between the metal nanostructures and 2D materials results in the plasmon damping. Therefore, the plasmon-exciton coupling is a plasmon damping channel in the hybrid system.

For the hybrid system of metal nanoparticles and 2D material, both charge transfer and resonance energy transfer mechanisms have been proposed to explain the interactions, such as the FWHM broadening when the GNRs interact with graphene[23]. But, conventional scattering linewidth broadening by itself cannot determine the relative contributions of these two plasmon damping pathways[24]. By the in *situ* single-particle spectroscopy, we can compare the scattering spectra of the same single GNRs on the glass, monolayer graphene, thin $h$BN flake, and the $h$BN/graphene, as shown in Figure 3d and 3e. The scattering intensity of the same GNRs on the $h$BN/graphene decreases obviously than that on the $h$BN due to the interaction between the GNR and graphene. The thin $h$BN flakes onto the graphene to prevent direct contact with the GNRs. The *h*BN as an intermediate medium can block the charge transfer between the GNR and the graphene. The wide bandgap *h*BN (here, its thickness is ~ 6 nm) would significantly inhibit the charge transfer between the GNR and graphene. Compared to the plasmon decay into metal electron-hole pairs followed by a charge transfer, resonant energy transfer is expected to be significant. Hence, the resonant energy transfer channel dominated the contributions of the plasmon decay in the GNR-graphene interaction rather than the charge transfer channels.

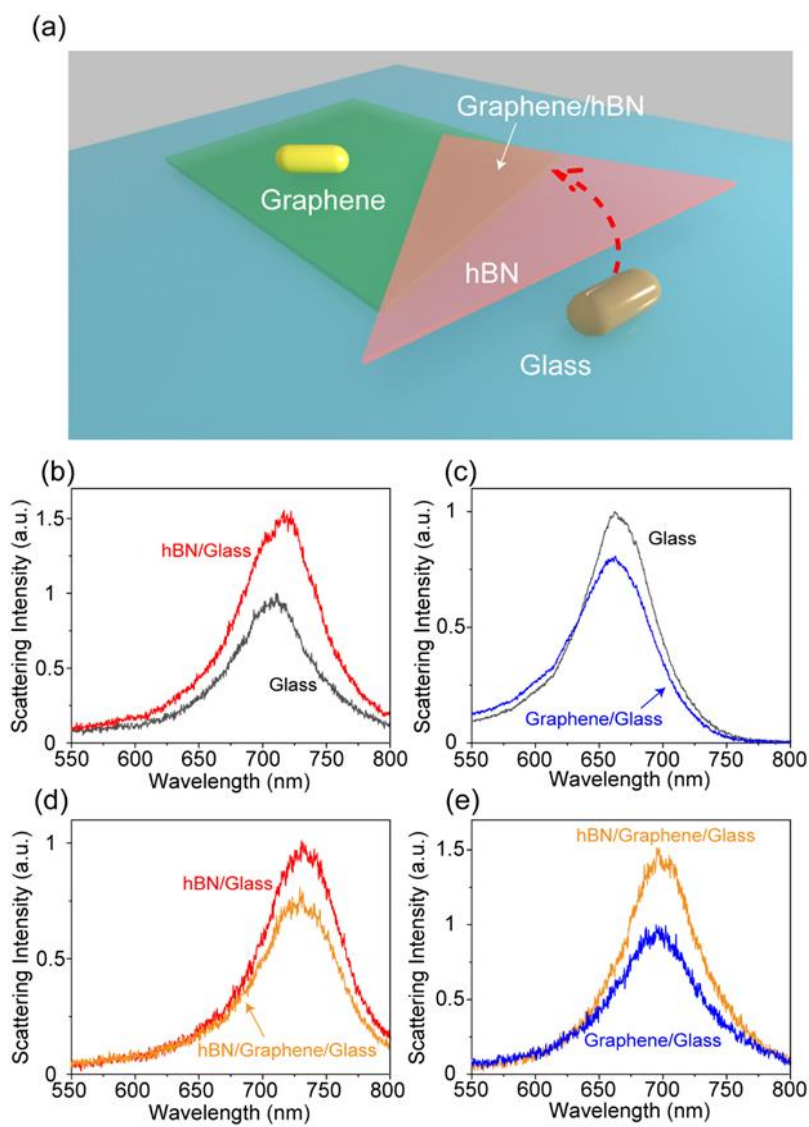

**Figure 3.** Scattering spectra of single GNR on the *h*BN/graphene layered structure. (a) Schematic illustration of a single GNR being manipulated with the AFM probe. (b) Scattering spectra of the same GNR on the glass (black) vs. on the *h*BN (red) flakes. (c) The same GNR which on the graphene (blue) vs. on the glass (black). (d) The same GNR on the *h*BN (red) vs.

on the $h$BN/graphene (orange). (e) The same GNR which on the $h$BN/graphene (orange) vs. on the graphene (blue).

Next, we investigate the influence of plasmon-exciton coupling strength on the plasmon damping. At first, we construct vertically stacked 2D materials $h$BN/WSe$_2$ heterojunction, as illustrated schematically in Figure 4a. The optical microscope image of the heterojunction is shown in Figure S1. The thin $h$BN layer can adjust the coupling strength between the GNR and the excitons in the WSe$_2$ because the presence of the $h$BN increases the distance between WSe$_2$ and the GNR. As shown in Figure 4b, when the GNR is being moved from the $h$BN to the $h$BN/WSe$_2$, the hybrid system exhibits a slight splitting feature, suffering a decrease in the scattering intensity by ~ 24%. We can rule out the charge transfer effect and interface contact effect in this situation since the GNR keeps in contact with the $h$BN surface. Hence, the coupling strength of plasmon-exciton interactions can influence the plasmon damping rate. Secondly, we show that the plasmon exciton coupling strength can be adjusted by the exciton transition dipole moment $\mu$ according to $g = \sqrt{N}\mu|E|/\hbar$, by tuning the layer number of the WSe$_2$ sheets[25-27]. Here, $\mu$ is exciton transition dipole moment, and $E$ is the electric field. It is expected that increasing the transition dipole moment can increase the plasmon exciton coupling strength. We obtained WSe$_2$ monolayer-bilayer homojunction sample through the mechanical exfoliation technique. As shown in Figure 4d, when the GNR being moved from monolayer WSe$_2$ to the bilayer WSe$_2$, the scattering intensity decrease by ~ 32% with deeper spectral dip and larger splitting, although the bilayer WSe$_2$ is indirect bandgap[28]. The increase of the number of excitons in the bilayer WSe$_2$ leads to the coupling strength increase, which suppresses the total plasmon oscillator strength. These results demonstrate the plasmon-exciton coupling strength can tune the plasmon damping rate.

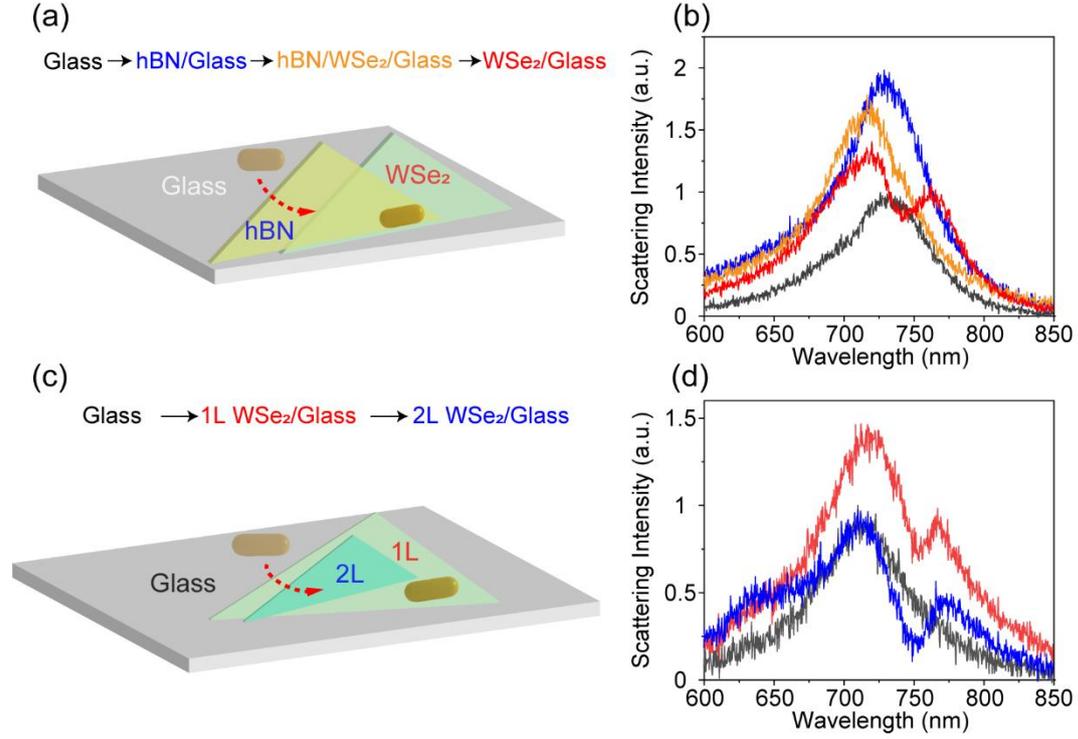

**Figure 4.** The coupling strength of plasmon-exciton interaction is controlled by separation and transition dipole moment. (a), (c) Schematic illustration of a GNR being moved from the glass onto 2D materials. (b) Scattering spectra of the same GNR on the glass (black), on the WSe$_2$ (red), on the WSe$_2$/$h$BN (orange), and the few layers $h$BN (blue). (d) Scattering spectra of the same GNR on the glass (black), on the monolayer WSe$_2$ (red), and bilayer WSe$_2$ (blue).

To understand the experiments intuitively, the two coupled oscillators model in the Hamiltonian representation provides a simple way to simulate plasmon-exciton interaction[29]. The coupled system is described by the following matrix equation:

$$\begin{pmatrix} \omega_1 - \omega - i\gamma_1 & g \\ g & \omega_2 - \omega - i\gamma_2 \end{pmatrix} \begin{pmatrix} x_1 \\ x_2 \end{pmatrix} = i \begin{pmatrix} f_1 \\ f_2 \end{pmatrix}$$

Where $x_1$ and $x_2$ are the oscillator amplitudes, $\omega_1$ and $\omega_2$ are the resonant frequencies, $g$ is the coupling strength between the oscillators, and $\gamma_1$, $\gamma_2$ are the damping coefficients, and $f_1$, $f_2$ are the external forces with the driving frequency $\omega$. We introduce the plasmon-exciton detuning as $\delta = \omega_{pl} - \omega_{ex}$. From the above equation, we can obtain the plasmon resonant scattering can be tuned by the plasmon-exciton coupling effect. The detuning and the coupling strength have a great influence on the shape of the scattering spectrum. In the fitting procedure, the plasmon linewidth was measured as the FWHM of the scattering spectrum of an uncoupled GNR ($\gamma_{pl} \approx 110\ meV$), and the exciton linewidth was determined from the PL spectra ($\omega_0 = 1.64\ eV$, $\gamma_0 = 50\ meV$). By changing the plasmon-exciton detuning or the coupling strength g, we can get the corresponding scattering spectra shown in Figure 5. The theoretical results show a similar conclusion about plasmon-exciton interaction in the plasmon damping as the experimental results.

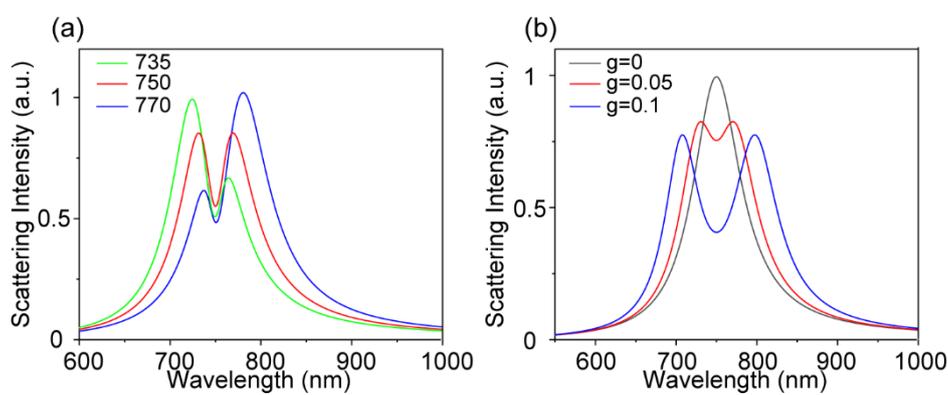

**Figure 5.** Scattering spectra calculated by coupled classical oscillator model. (a) Calculated scattering spectra for different plasmon-exciton detuning and (b) for different coupling strengths.

Interestingly, the scattering intensity increases significantly after the GNR being moved onto the TMDCs and $h$BN layers from the glass surface when the plasmon-exciton detuning is large (e.g., over an FWHM of the GNR's plasmon resonance). To confirm scattering enhancement, we measured the transmission spectra through the glass slide with different thickness $h$BN sheets. As shown in Figure S3, the transmittances are almost the same, which confirms that we can rule out the scattering enhancement caused by enhancing the reflection of the $h$BN layers. The influence of radiative directivity on the light collection efficiency is also negligible. As shown in Figure S4, the presence of the $h$BN and WSe$_2$ (i.e., increasing of the effective index for bottom half-space) increases the electromagnetic flux toward the glass substrate by less than 2% compared with that of the GNR on the glass surface. For graphene, the downward collection ratio decreases about 0.4% compared to glass. We performed the finite difference time domain (FDTD) method to simulate the interaction between the GNR and 2D materials. Figure S5 shows the calculated scattering spectra of the different GNRs on the glass with or without the monolayer $WSe_2$, $h$BN sheet, and graphene. The numerical results can explain the spectral shift and split due to the classic electromagnetic interaction[22]. But the variations of scattering intensity by the FDTD simulation are inconsistent with the experimental results. The dielectric interaction between the GNRs and 2D materials is limit to explain the interfacial interaction effect[22].

Furthermore, a bulk material effect on the plasmon damping has been demonstrated by comparison the linewidth of Ag clusters in $SiO_2$ and $Al_2O_3$[30]. Here, we find that the first contact layer between the GNR and 2D materials is critical for the interfacial interaction effect. The increase of scattering intensity is about 70% after the GNR is moved onto the $WSe_2$ from the glass surface when the plasmon-exciton detuning is large, which is about 130 % after being moved onto thin $h$BN flakes, as shown in Figure 4. We also record the scattering of the same GNR on the different thicknesses of $h$BN, as shown in Figure S6. As the thickness increases from 10 to 30 nm, the effective refractive index feeling by the GNR increases. So, the plasmon resonance energy shifts to lower energy. However, the scattering intensity remains constant rather than increasing scale with layer thickness. Another clue is that the scattering intensity increasing on the 1L and 2L $WSe_2$ compared to that on the glass is comparable when the plasmon-exciton detuning is large (the data not shown here). Hence, the interfacial interaction effect contributing to the plasmon damping is mainly induced by the first contact layer. These interfacial interactions may have various origins, owing to the electron scattering interface by such as defects and imperfections at the interface[31], induced surface dipoles in the metal[9], the strain of the metal surface[32], or different orientations of the adsorbate on the atomistic scale. Nevertheless, these results imply that decoration or modification of the first contact layer of the metal particle surface is critical to control the plasmonic damping or resonant quality factor.

**Conclusion**

The interactions between a single GNR and 2D materials were investigated using the single-particle spectroscopy method by comparing the same particle's scattering intensity and linewidth before and after coupling. This approach allows us to circumvent the drawback of the conventional single-particle method to evaluate the plasmon damping rate by the linewidth of scattering spectral. We demonstrate that the interface interaction between the GNRs and 2D materials greatly influences plasmon decay. We find out that the plasmon-exciton coupling in the GNR-WSe$_2$ hybrid would induce plasmon resonance damping, and the damping rates are dependent on the detuning and coupling strength. We also confirm that the contribution of resonant energy transfer dominates the plasmon damping rather than the electron transfer mechanism in the GNR-graphene hybrid. The contact layer between the GNR and 2D materials contributes mainly to the interfacial plasmon damping effect. Further controlled experiments are still needed to reveal the physical origin of the increase of scattering intensity. Overall, these systematic studies pave the way to a predictive understanding of plasmon decay channels through the interface interaction of metal nanoparticles and 2D materials.

**Methods**

**Sample Preparation.**

WSe$_2$ monolayer samples were grown on the sapphire substrate through a chemical vapor deposition (CVD) technique (Carbon Six Ltd.). As-grown graphene film on Cu foil was purchased from the Carbon Six Ltd. Gold nanorods were synthesized using the seed-mediated method. Then, the glass coverslips were cleaned by ultrasonic washing, then treated by oxygen plasma for 5 min, and were immersed in MPTMS alcohol solution (5%, v/v) for 10 min. Then, the GNRs solution is drop-casted on this

MPTMS-functionalized surface. These are spaced far enough apart (>5 μm) to observe them individually. The CVD-grown monolayer $WSe_2$ was transferred onto the glass immobilized GNR[33].

*h*BN can be obtained via the mechanical exfoliation technique, where blue tape is used to cleave the *h*BN crystals till they are sufficiently thinned down. The thinned-down *h*BN microflakes, comprising few-layered and multilayered materials, are transferred onto polydimethylsiloxane (PDMS) substrates. Atomic force microscope (AFM) measurements can obtain the thickness of the *h*BN. Subsequently, they are transferred onto glass substrates with immobilized GNRs using the dry transfer technique with a micromanipulator and the aid of a heating pad. The transfer method of the monolayer graphene was the wet transfer [34].

The graphene/*h*BN sample can be obtained by transferring the *h*BN, which was mechanical exfoliation to glass coverslip overlapping with monolayer graphene with the same method as before. The $WSe_2$/*h*BN sample can be obtained by transferring the *h*BN to the glass coverslip overlapping with monolayer $WSe_2$ (CVD) by the dry transfer technique with a micromanipulator[35]. The $WSe_2$ flakes with different layers were mechanically exfoliated onto polydimethylsiloxane stamps using the same method with the *h*BN and transferred on the glass immobilized GNR. The CCD contrast was used to identify the different layers of $WSe_2$. Photoluminescence was carried out to confirm the monolayer and bilayer nature.

**Experimental Section**

**Single Particle Spectroscopy.** Scattering spectra of single gold nanorods were acquired with a scanning probe microscope (NTEGRA Spectra, NT-MDT, Russia), which was developed to combine white light dark-field scattering and photoluminescence and atomic force microscopy. A detailed description of this optical setup is given by Zhang et al. [18-19] The scattering spectra of nanoparticles were measured using a white light total internal reflection scattering method based on a high numerical aperture oil-immersion objective lens (NA 1.49, 60 ×TIRF, Olympus, Japan). A collimated white light beam was approximately focused at the objective's back focal plane, and the scattering signal was collected by the same objective lens and directed into a spectrometer with a cooled CCD (iDdus, Andor).

**Transmittance Spectroscopy.** A spectrometer in the visible range analyzed the light passing through the glass and different area $h$BN flakes. The transmittance spectra microscope is operated in epi-illumination mode. The source using a small halogen tungsten lamp (Ocean Insight) was entered into the beam path by the optical fiber. Then, the light spot through a long working distance (infinity corrected) 100× microscope objective and the size of the light spot on the sample is about 1 to 2 um. An XY stage allows to move the sample position, and the bright spots in Figure S3a are spectral collection points. The transmittance spectra can be collected by a fiber-coupled CCD spectrometer (Ocean Insight) under the sample holder.

ASSOCIATED CONTENT

**Supporting Information**.

The Supporting Information is available free of charge online.

Experimental details; supplementary optical image and AFM images for sample; PL images of 2D material; transmission spectra for $h$BN; FDTD simulate details and result; the scattering spectra of GNR on the different layers $h$BN.

AUTHOR INFORMATION

**Corresponding Author**

Email: guowei.lu@pku.edu.cn

**Author Contributions**

G.L. conceived the original ideas presented in this work and constructed the experimental setup. L.Y. performed the experiments. L.Y. and W.Z. developed and performed the numerical simulations, the theoretical modeling, and the interpretation of data. Y.G. and L.T. contributed to the experimental realization and interpretation of the results. L.Y. and G.L. jointly wrote the manuscript. All the authors discussed the results and wrote the manuscript.

**Notes**

The authors declare that they have no competing interests.

ACKNOWLEDGMENTS

This work was supported by the National Key Research and Development Program of China (Grant No. 2018YFB2200401), the Guangdong Major Project of Basic and Applied Basic Research (Grant No. 2020B0301030009), and the National Natural Science Foundation of China (Grant Nos. 91950111, 61521004, and 11527901).